\documentclass[aps,prb,twocolumn,amsmath,amssymb,floats,superscriptaddress]{revtex4-1}
\usepackage[dvips]{graphicx}
\usepackage{color}
\usepackage{float}
\usepackage{epsfig}
\usepackage{epstopdf}
\usepackage{multirow}
\usepackage[none]{hyphenat}
\usepackage{dcolumn}
\definecolor{url}{RGB}{0,20,160}
\usepackage[colorlinks=true,linkcolor=blue,citecolor=blue,urlcolor=url]{hyperref}
\usepackage[usenames,dvipsnames,svgnames,table]{xcolor}

\newcommand{\be}{\begin{equation}}
\newcommand{\ee}{\end{equation}}
\newcommand{\bea}{\begin{eqnarray}}
\newcommand{\eea}{\end{eqnarray}}
\newcommand{\ba}{\begin{eqnarray*}}
\newcommand{\ea}{\end{eqnarray*}}

\newcommand{\up}{\uparrow}
\newcommand{\down}{\downarrow}

\newcommand{\eqn}[1]{(\ref{#1})}

\newcommand{\ket}[1]{\mid #1 \rangle}
\begin{document}
 
\title{Nanoscale orbital excitations and the infrared spectrum of a molecular Mott insulator: A15-Cs$_{3}$C$_{60}$}

\author{S.~Shahab Naghavi}
\affiliation{Department of  Materials Science and Engineering,  Northwestern University, Evanston, Illinois 60208, USA}
\author{Michele Fabrizio}
\affiliation{International School for Advanced Studies (SISSA), and CNR-IOM Democritos National Simulation Center, Via Bonomea 265, I-34136 Trieste, Italy }
\author{Tao Qin}
\affiliation{International School for Advanced Studies (SISSA), and CNR-IOM Democritos National Simulation Center, Via Bonomea 265, I-34136 Trieste, Italy }
\affiliation{Institut f\"ur Theoretische Physik, Goethe-Universität, 60438 Frankfurt am Main, Germany}
\author{Erio Tosatti}
\affiliation{International School for Advanced Studies (SISSA), and CNR-IOM Democritos National Simulation Center, Via Bonomea 265, I-34136 Trieste, Italy }
\affiliation{International Centre for Theoretical Physics (ICTP), Strada Costiera 11, I-34151 Trieste, Italy 
}

%\email[]{E-mail: tosatti@sissa.it}
%%%%%%%%%%%%%%%%%%%%%%%%%%%%%%%%%%%%%%%%%%%%%%%%%%%%%%%%%%%%%%%%%%%%%%%%%%%%%
%\date{\today}
%%%%%%%%%%%%%%%%%%%%%%%%%%%%%%%%%%%%%%%%%%%%%%%%%%%%%%%%%%%%%%%%%%%%%%%%%%%%%
\begin{abstract} 
The quantum physics of ions and  electrons behind low-energy spectra of strongly
correlated {\it  molecular} conductors,  superconductors and Mott  insulators is
poorly  known, yet  fascinating especially  in orbitally  degenerate cases.  The
fulleride insulator  Cs$_{3}$C$_{60}$ (A15), one such  system, exhibits infrared
(IR) spectra  with low  temperature peak features  and splittings  suggestive of
static  Jahn-Teller distortions  with  breakdown  of  orbital symmetry  in  the
molecular site.  That is puzzling, for there is no detectable static distortion,
and because  the features  and splittings disappear  upon modest  heating, which
they should  not. Taking  advantage of the  Mott-induced collapse  of electronic
wavefunctions  from  lattice-extended  to  nanoscale localized  inside  a  caged
molecular  site, we  show that  unbroken spin  and orbital  symmetry of  the ion
multiplets  explains  the  IR   spectrum  without  adjustable  parameters.  This
demonstrates the  importance of a  fully quantum treatment of  nuclear positions
and orbital momenta in the Mott  insulator sites, dynamically but not statically
distorted. The observed  demise of these features with  temperature is explained
by  the thermal  population  of a  multiplet term  whose  nuclear positions  are
essentially undistorted,  but whose energy  is very  low-lying. That term  is in
facts  a  scaled-down  orbital  excitation  analogous  to  that  of  other  Mott
insulators,  with the  same spin  $1/2$ as  the ground  state, but  with a  larger
orbital moment of two  instead of one.
\end{abstract}
%%%%%%%%%%%%%%%%%%%%%%%%%%%%%%%%%%%%%%%%%%%%%%%%%%%%%%%%%%%%%%%%%%%%%%%%%%%%%%
\maketitle
\section{Introduction}

Crucial  in  atomic  and  molecular  physics, orbital  degrees  of  freedom  are
generally quenched  by crystal  fields and  by large  electron hopping  rates of
uncorrelated  solids, where  they  are  seldom relevant  to  the electronic  and
vibrational structure.  That is no longer the case in Mott insulators, where the
electrons  are  localized by  repulsive  correlations,  and their  wavefunctions
revert from  band states back to  effectively atomic or molecular  states, whose
electronic,  orbital,  spin,  and  nuclear  degrees  of  freedom  are  generally
entangled  in the  crystalline  environment.   The role  of  orbital degrees  of
freedom    in    particular,    long    appreciated    in    transition    metal
compounds\,\cite{Goodenough1971,Kugel-Khomskii82,  Schlappa2012, Castellani1978,
Ishihara1997,VandenBrink2001,Khaliullin2000} is less explored in {\it molecular}
Mott insulators (MMIs), where site symmetry is often low, and where a variety of
intra-molecular interactions interfere.  A fresh  case study is highly desirable
in  these  systems,  particularly  in  connection  with  possible  spectroscopic
manifestations  of  orbital  variables   and  orbital  excitations.   Especially
appealing are MMIs where a high symmetry of the molecular site and the resulting
orbital degeneracies, quenched in the  uncorrelated band state but unquenched in
the  Mott state,  demands an  explicitly  joint role  for the  spin and  orbital
degrees of  freedom of both  electrons and  ions. Particularly appealing  is the
opportunity, typical  of Mott  insulators and generally  only exploited  by DMFT
approaches,  to  access  the  main   physical  phenomena  through  theories  and
calculation  restricted to  a nanoscale  size molecular  site, endowed  with the
site's  full orbital  symmetry, but  with a  reduced scale  intra-site molecular
interactions as compared to well studied atomic cases. Besides others, one major
difference between them is that the nanoscale size of the molecular site shrinks
the Hund's rule exchange energy $J_{\rm H}$ one to two orders of magnitude below
the   1--2\,eV   scale  typical   of   an   atomic   Mott  insulator   such   as
Sr$_{2}$CuO$_{3}$.~\cite{Schlappa2012}  This  in   turn  causes  the  intra-site
electronic multiplets including orbital excitations to drop and become entangled
with  the  lower  energy  nuclear  vibrations,  unlike  e.g.,  transition  metal
compounds  where the  orbital  excitations,  way higher  in  energy, are  purely
electronic in character.~\cite{Schlappa2012}

The  prototypical  MMI  compound  which  embodies these  novel  aspects  is  the
over-expanded  fulleride  A15-Cs$_{3}$C$_{60}$,  a $S\!=\!1/2$  Mott  insulator,
where a striking superconducting state  with $T_{\rm c}\,{\sim}\,38$\,K also emerges
under         pressure         from        the         parent         insulating
state.~\cite{ganin08,Takabayashi09,Ihara10,Ganin10,kamaras2014,Zadik2015,Baldassarre2015}
Pressure here  mostly acts  by reducing  the ratio  $U/W$ of  the intramolecular
Coulomb repulsion  $U$ over  the intermolecular electron  bandwidth $W$  below a
critical value of order 1--2, turning the MMI into a half-filled three-band metal
state  typical  of  other  non-expanded  alkali  fullerides.~\cite{Gunnarsson97}
Thanks to  many studies the  spin structure and spin  excitations of the  MMI at
zero              pressure              have              been              well
characterized.~\cite{Takabayashi09,Jeglic09,Ihara10,Wzietek14,chibotaru05,
iwahara13, iwahara15} Their  orbital counterparts are on the  contrary much more
obscure, making this compound an ideal laboratory for our purposes.

We focus here  on the pure, dynamical MMI state,  with full unbroken structural,
magnetic  and   orbital  site  symmetry---a   state  which  as  in   other  Mott
insulators,~\cite{Georges1996} is attained in thermal equilibrium just above the
antiferromagnetic N\'eel  temperature, $T_{\rm  N}\,{\sim}\,46\,K$ at  zero pressure.
In that state, A15-Cs$_{3}$C$_{60}$ has a lattice of space group ${Pm\bar{3}n}$,
where  Mott localized  electrons hop  between effective  C$_{60}^{3-}$ sites  of
cubic  symmetry,  a  situation  qualitatively   described  by  a  high  symmetry
three-band Hubbard  model.~\cite{Gunnarssson96,Capone02, Capone09,  Nomura16} In
the  Mott state,  intersite  charge  fluctuations are  suppressed  by the  large
on-site Coulomb  repulsion $U\,{\gtrsim}\,1$~eV,~\cite{Gunnarsson97,Capone02} and
the  system can  be assimilated  to weakly  coupled effective  molecular anions,
where  interactions between  sites  can be  approximately  neglected.  Thus  the
orbital  and  spin-dependent quantum  mechanical  energy  spectrum of  a  single
effective  C$_{60}^{3-}$  ion,  once  duly   immersed  in  a  crystal  field  of
appropriate strength  and cubic  T$_d$ symmetry, as  sketched in  Fig.~1, should
provide a  good representation  for the  MMI state  and its  excitation spectrum
above  $T_{\rm N}$.   With  three  electrons in  three  $t_{\rm 1u}$  ($p$\,-like)
orbitals, the effective ion site multiplet  consists of, at frozen nuclei, three
electronic terms, a high spin  quartet $^{4}A$ ($L\!=\!0$, $S\!=\!3/2$), and two
low  spin  doublets  $^{2}H$  ($L\!=\!2$,  $S\!=\!1/2$)  and  $^2T$  ($L\!=\!1$,
$\!S=\!1/2$).   On  account of  Hund's  first  rule,  which dictates  an  energy
$-J_{\rm   H}\,[2S(S+1)+L(L+1)/2]$,~\cite{Capone09}  the   spin   3/2  term   at
$E=-(15/2)J_H$ should at frozen nuclei be the ground state, lying lower than the
two    spin   1/2    states,    $E=-(9/2)J_{\rm    H}$,   $E=-(5/2)J_{\rm    H}$
respectively.~\cite{Negri1992} However, as  was anticipated, the intra-molecular
exchange  $J_{\rm H}\,{\sim}\,50$\,meV~\cite{Capone09}  is small,  comparable with
vibrational   energies  ($30$--$150$\,meV)   invalidating  the   frozen  nuclei
approximation, and causing a dynamical Jahn-Teller entanglement with vibrational
states.  That  leads to a  drastic reversal of  these three states,  now endowed
with      joint      electron-nuclear     character.       Despite      existing
treatments~\cite{Auerbach1994,obrien96,   chibotaru05,   iwahara13,   iwahara15}
neither the  specific role of orbital  degrees of freedom nor  the spectroscopic
consequences of  this entanglement appear  to have been adequately  addressed so
far.   The IR  spectra in  particular  require specific  first principles  based
calculations  for   each  dynamical  multiplet  term   characterising  the  Mott
insulator, an agenda  not implemented until now, but indispensable  here as well
as in future applications.
%
%%%%%%%%%%%%%%%%%%%%%%%%%%%%%%%%%%%%%%%%%%%%%%%%%%%%%%%%%%%%%%%%%%%%%%%%%%%%%%%%%
\begin{figure}[htp!]
	\centering
	\includegraphics[width=0.95\linewidth]{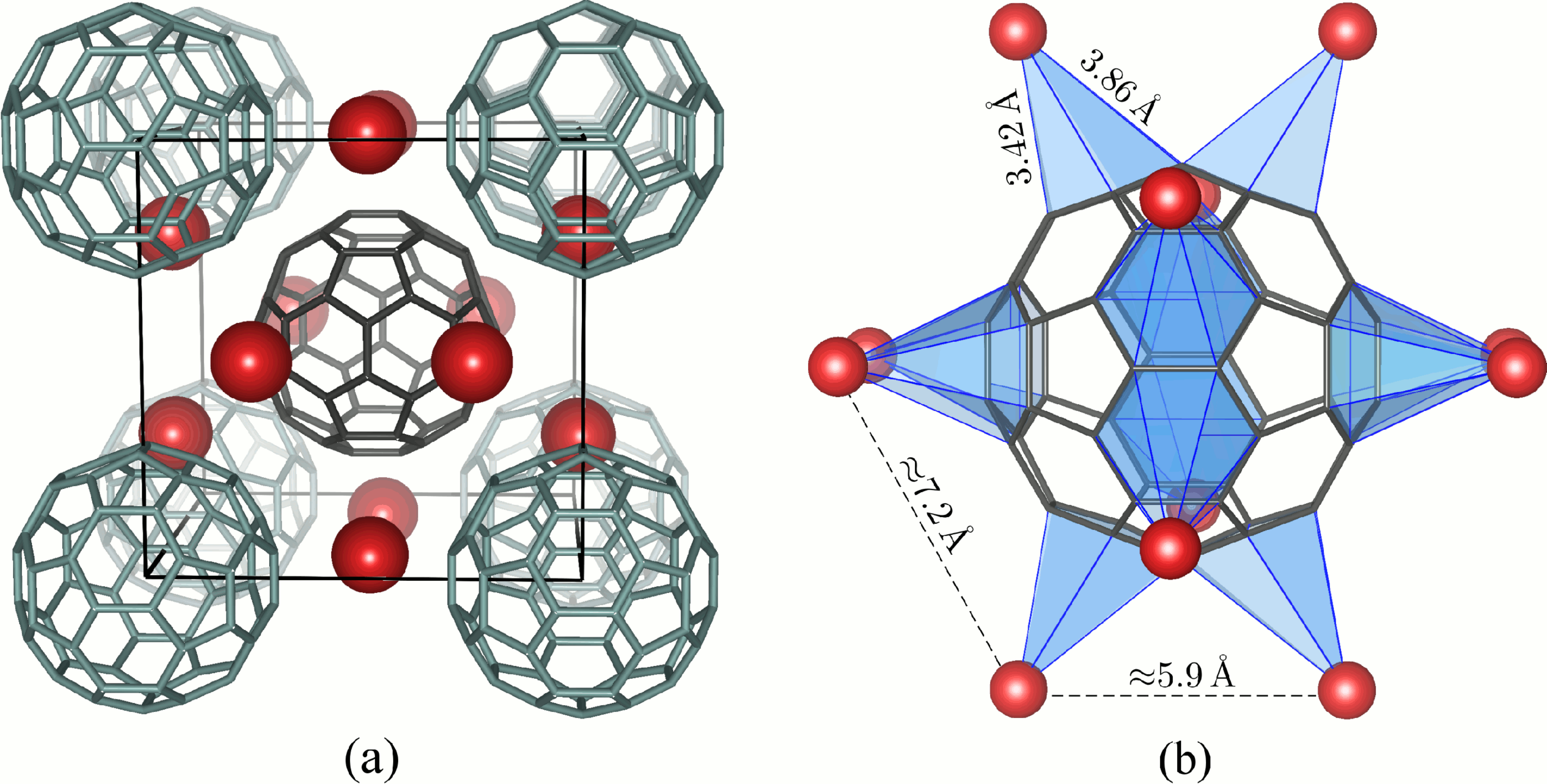}
	\caption{(a)              Crystal              structure              of
A15--Cs$_3$C$_{60}$.~\cite{Takabayashi09}   (b)    Assumed   C$_{60}^{3-}$   ion
geometry.  The crystal field is mimicked  by 12 point charges $q$=0.25\,$|e|$ at
the Cs nuclear positions in Ganin et\,al.~\cite{ganin08}}
\end{figure}
%%%%%%%%%%%%%%%%%%%%%%%%%%%%%%%%%%%%%%%%%%%%%%%%%%%%%%%%%%%%%%%%%%%%%%%%%%%%%%%%%
%
The MMI  spectrum is  calculated as that  of an uncoupled  ion in  the nanoscale
crystal field model  as on Fig.~1 (parameters given in  Methods), in four steps.
First, treating carbon nuclear coordinates as static variables, state-of-the-art
density  functional theory  (DFT) electronic  structure and  molecular vibration
calculations are carried  out yielding the total energy,  the optimal distortion
away from T$_{\rm d}$, and the full vibrational spectrum of the three statically
distorted  adiabatic states  of spin  3/2  and 1/2  $\alpha$, $\beta$,  $\gamma$
realized by  three electrons  in the initially  $t_{\rm 1u}$  molecular orbital.
Second,  for each  of these  adiabatic and  distorted states  the IR  absorption
spectra are calculated, again by accurate  DFT methods. Third, the carbon nuclei
in these statically distorted but unphysical states are allowed to delocalize by
weak  tunneling  so as  to  give  rise  to  three fully  T$_{\rm  d}$-symmetric
dynamical joint  multiplet terms  $^{4}A$, $^{2}H$, and  $^2T$ of  electrons and
ions, with  different energies  from the  static DFT  states. Fourth  and final,
results for the IR absorption spectra  corresponding to each multiplet state are
derived  from  the  previously  calculated adiabatic  spectra.  Comparison  with
experiments clarifies the  significance of IR peaks and  splittings, whereas the
overall temperature  dependence reveals the  thermal population of  an important
low lying orbital excitation.

\section{Methods}
Near the  C$_{60}^{3-}$ ion the  crystal field is  mimicked by 12  point charges
$q$=0.25\,$|e|$ at the  Cs nuclear positions in  Ganin et\,al.,~\cite{ganin08} a
setup  which  also  averts  self-ionization, yielding  a  ionization  energy  of
approximately $2$\,eV  similar to that  of metallic fullerides.   Spin polarized
DFT  calculations  are carried  out  with  the  NWCHEM  code. The  B3LYP  hybrid
functional~\cite{Becke}   polarization  (DZVP)   basis-set   of  15   contracted
functions,  providing  excellent  DFT  treatments of  carbon  --  especially  of
exchange, here of vital importance.  Accurate relaxation of C-atom positions and
total energy  minimization at  $T$=0 gave  rise to  distortions, large  in state
$\gamma$, nil in $\alpha$ and small but  nonzero in $\beta$, owing to a residual
but real  energy gain caused  by splitting  majority from minority  spin levels.
{\sl Ab-initio}  calculation of  the 174  vibrational frequencies  $\omega_i$ is
carried out  for fully relaxed  states $\alpha$,  $\beta$, and $\gamma$  and the
resulting zero-point  energies E$_{\alpha}$, E$_{\beta}$, E$_{\gamma}$  given in
the  text.  The  IR absorption  spectra of  states $\alpha,  \beta, \gamma$  are
calculated  by evaluating  dipole moments  by numerical  differentiation of  the
gradient    at     the    equilibrium    geometries,    as     implemented    in
\texttt{NWCHEM}.~\cite{nwchem}

\section{Results and discussion}

\subsection{Adiabatic states}

We  find by  state-of-the-art DFT  calculations (details  given in  Methods) the
lowest   energy  electronic   structure  and   static  nuclear   coordinates  of
C$_{60}^{3-}$  in   the  cubic  environment   of  Fig.~1  for   three  different
configurations, obtained by the following  occupancies of three orbitals denoted
as  ($x$,$y$,$z$):   $\ket{\!\alpha}$:  occupancies  $n_{x   \up}\!=\!1$,  $n_{y
\up}\!=\!1$,  $n_{z   \up}\!=\!1$  ,   optimal  nuclear  symmetry   T$_{\rm  d}$
$S_z\!=\!3/2$;   $\ket{\!\beta}$:    occupancies   $n_{x    \up}\!=\!1$,   $n_{y
\down}\!=\!1$,  $n_{z  \up}\!=\!1$,  optimal   nuclear  symmetry  D$_{\rm  5d}$,
$S_z\!=\!1/2$;  $\ket{\!\gamma}$:  occupancies  $n_{x\up}\!=\!n_{x\down}\!=\!1$,
$n_{z  \up}\!=\!1$,  optimal  nuclear   symmetry  D$_{\rm  2h}$,  $S_z\!=\!1/2$.
Structurally, state $\ket{\!\alpha}$ is undistorted, 24 of its carbon atoms at a
distance from  the center  $d=d_0=12.7175$, 24  at 12.7095 A,  12 at  12.6997 A.
State  $\ket{\!\gamma}$ is  on  the contrary  heavily  Jahn-Teller D$_{\rm  2h}$
distorted,  with  $\sqrt{(1-d/d_0)^2}   \sim  10^{-2}$.   State  $\ket{\!\beta}$
finally,  exhibits  a  very  small  D$_{\rm 5d}$  distortion,  entirely  due  to
electron-electron interactions, with $\sqrt{(1-d/d_0)^2} \sim 10^{-4}$. For each
optimal   state  $\ket{\!\alpha}$,   $\ket{\!\beta},  \ket{\!\gamma}$   we  then
calculate the full vibrational spectrum  and from that the respective zero-point
energy  correction,  ending  up  with   the  following  total  energy  sequence:
$E_{\alpha}-E_{\gamma}=    66.0$\,meV    and    $\text{E}_\beta-\text{E}_\gamma=
27.1~$\,meV.

\subsection{IR spectra of adiabatic states}

In each  adiabatic state $\ket{\!\alpha}$, $\ket{\!\beta}$,  $\ket{\!\gamma}$ we
calculate, as  detailed in  Methods, the {\sl  ab-initio} IR  absorption spectra
$I_\alpha$, $I_\beta$ and  $I_\gamma$ shown in Fig.~2 in comparison  with the IR
spectrum calculated for neutral C$_{60}$ in same geometry (panel (a)).

%%%%%%%%%%%%%%%%%%%%%%%%%%%%%%%%%%%%%%%%%%%%%%%%%%%%%%%%%%%%%%%%%%%%%%%%%%%%%%%%
\begin{figure}[htp!]
	\centering
	\includegraphics[width=0.95\linewidth]{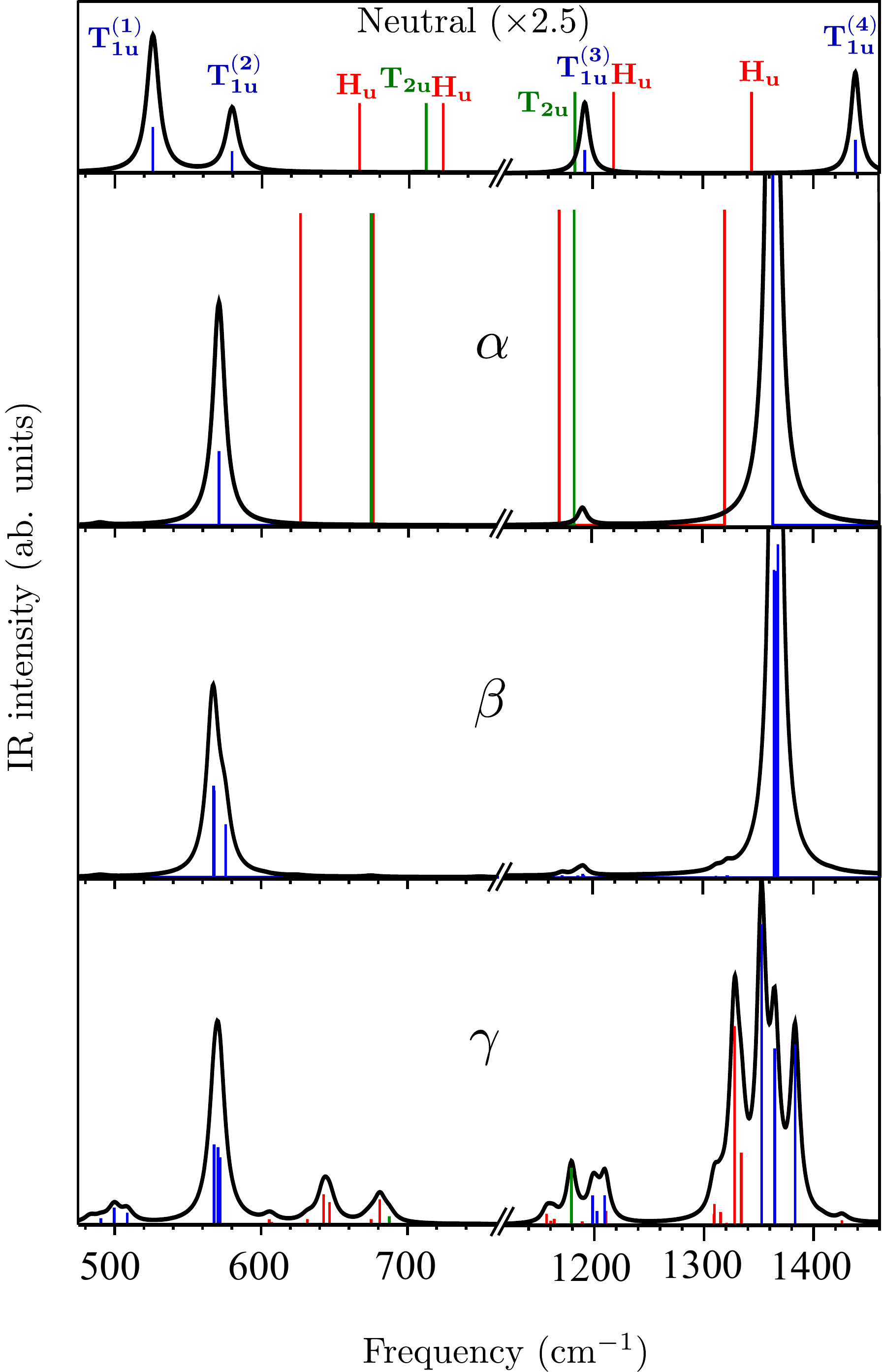}
	\caption{(a)--(d):  Calculated  IR  spectra  of  neutral
C$_{60}$  and of  C$_{60}^{3-}$  in DFT  statically  optimized states  $\alpha$,
$\beta$ and $\gamma$.  The neutral case scale is multiplied by 6.  Red and green
lines indicate  optically forbidden  modes, becoming  partly allowed  in heavily
distorted state $\gamma$.}
\end{figure}
%
%%%%%%%%%%%%%%%%%%%%%%%%%%%%%%%%%%%%%%%%%%%%%%%%%%%%%%%%%%%%%%%%%%%%%%%%%%%%%%%%

Neutral C$_{60}$ possesses four IR active $T_{\rm 1u}(n)$ modes, $n=1,...4$.  In
the crystal-field caged ion, IR peaks  are red shifted relative to their neutral
counterparts -- especially T$_{\rm 1u}(4)$.   Moreover, the T$_{\rm 1u}(1)$ peak
amplitude  near $520\,\text{cm}^{-1}$  vanishes  in the  ion  while the  T$_{\rm
1u}(3)$  peak  at  about  1200\,cm$^{-1}$ weakens.  Two  additional  differences
characterize the heavily  distorted state $\ket{\!\gamma}$. First,  the IR modes
are split and  modified due to nonlinearities caused by  the nuclear distortion,
each  T$_{\rm 1u}$  mode  splitting into  B$_{\rm  u}$+A$_{\rm u}$+B$_{\rm  u}$.
Second, other {\sl ungerade} modes such as  $H_u$ and $T_{2u}$, IR silent in the
neutral undistorted  molecule, develop  components with  an electric  dipole, IR
active in this state. Conversely, in state $\beta$ (negligibly small distortion)
and in  $\alpha$ (zero distortion) there  are no visible mode  splittings and no
new lines, so  the IR spectrum is  much simpler and closer  to neutral C$_{60}$.
These  calculated spectra  are  not final  however,  the calculations  requiring
quantum delocalization  of nuclei  before experimental comparison,  as explained
below.

\subsection{Fully symmetric multiplet states}
The adiabatic states $\ket{\!\alpha}$,  $\ket{\!\beta}$, $\ket{\!\gamma}$ do not
correctly describe  MMI multiplets ,  because by construction  they artificially
break  both  spatial  and  spin  rotational invariance---they  generally  are
symmetry-broken, as well as spin contaminated. The real dynamical MMI state must
have fully orbital  and spin rotational symmetry, namely $^2T$,  $^4A$ and $^2H$
now again  within T$_{\rm d}$  ion, but with  different IR properties,  which we
presently    calculate.     Consider     for    instance    $S_z{=}1/2$    state
$|\gamma\rangle\,{=}\,c^\dagger_{z\up}c^\dagger_{x\up}c^\dagger_{x\down}|0\rangle\times
|\Phi\rangle$.   The   vibrational  wavefunction   $\Phi$  can  be   written  as
$\Phi\,{=}\,\Phi_\text{e}\,{+}\,\Phi_\text{o}$, the  sum of two  components that
differ by parity, even (e) or odd (o), of the number of vibrational modes of the
undistorted  C$_{60}^{3-}$.  The  approximate  expression  of  the  rotationally
symmetric doublet state $^2T$ is

%%%%%%%%%%%%%%%%%%%%%%%%%%%%%%%%%%%%
\begin{eqnarray}
|^2T\rangle &\sim& c^\dagger_{z \sigma}\,\big(c^\dagger_{x \uparrow}c^\dagger_{x \downarrow}
+ c^\dagger_{y \uparrow}c^\dagger_{y \downarrow}\big)\,|\Phi_\text{e}\rangle \nonumber\\
&& + 
c^\dagger_{z \sigma}\,\big(c^\dagger_{x \uparrow}c^\dagger_{x \downarrow}
- c^\dagger_{y \uparrow}c^\dagger_{y \downarrow}\big)\,|\Phi_\text{o}\rangle,\label{2T1u}
\end{eqnarray}
%%%%%%%%%%%%%%%%%%%%%%%%%%%%%%%%%%%%
Its orthogonal doublet partner is instead
%%%%%%%%%%%%%%%%%%%%%%%%%%%%%%%%%%%%
\begin{eqnarray}
|^2H\rangle &\sim& c^\dagger_{z \sigma}\,\big(c^\dagger_{x \uparrow}c^\dagger_{x \downarrow}
+ c^\dagger_{y \uparrow}c^\dagger_{y \downarrow}\big)\,|\Phi_\text{o}\rangle \nonumber\\
&& + 
c^\dagger_{z \sigma}\,\big(c^\dagger_{x \uparrow}c^\dagger_{x \downarrow}
- c^\dagger_{y \uparrow}c^\dagger_{y \downarrow}\big)\,|\Phi_\text{e}\rangle,\label{2Hu}
\end{eqnarray}
%%%%%%%%%%%%%%%%%%%%%%%%%%%%%%%%%%%%
%
It thus follows that   
\be
|\gamma\rangle \sim  \sqrt{\frac{1}{2}}\; \Big[|^2T\,\rangle + |^2H\,\rangle\Big]. \label{gamma-rough}
\ee
By a similar argument we infer that 
\be
|\beta\rangle = c^\dagger_{x\up} c^\dagger_{y\down}c^\dagger_{z\up} |0\rangle \times |\Phi'\rangle
\sim \sqrt{\frac{1}{3}}\; |^4A\rangle + \sqrt{\frac{2}{3}}\; |^2H\rangle,\label{beta-rough}
\ee
while $|\alpha\rangle = |^4A\rangle$. 
Direct support to Eqs. \eqn{gamma-rough} and \eqn{beta-rough} comes from calculating by DFT the total spin 
expectation values, which we find to be ${<\alpha|\rm  S^{2}|\alpha>}  \simeq   15/4$,  ${<\gamma|\rm
	S^{2}|\gamma>} \simeq  3/4$, and  ${<\beta|\rm S^{2}|\beta>}= 1.77 \simeq  7/4 $ exactly as predicted
by these equations. Inverting the transformation, 
\begin{eqnarray}
E_{\,^4A}&=&E_{\alpha}, \nonumber\\
E_{\,^2H}&=&\frac{3}{2}E_{\beta} - \frac{1}{2}E_{\alpha}, \label{energies} \\
E_{\,^2T}&=&2E_{\gamma}-\frac{3}{2}E_{\beta}+ \frac{1}{2} E_\alpha, \nonumber
\end{eqnarray}
where $ E_{\,^4A} = E_{\alpha}$ is now set conventionally to zero. Inserting the
calculated values of $E_{\gamma}$ and  $E_{\beta}$ we obtain the final multiplet
energies, namely $E_{\,^2H}= -58.3$ meV, $E_{\,^2T}= -73.7$ meV $ E_{\,^4A}=0 $.
The 1.6 $\mu_B$ moment observed by NMR in A15 Cs$_{3}$C$_{60}$ above $T_{\rm N}$
agrees  well  with  1.73  of  our $^2T$  ground  state.~\cite{Jeglic09}  A  spin
excitation   gap  of   order  800\,K   well   established  here   as  in   other
fullerides~\cite{brouet01,Jeglic09}  is in  agreement with  the large  $^4A-^2T$
calculated gap.

Now, as  anticipated, the  $^2H$ state is  found a mere  $\sim$ 15  $\pm$ 6\,meV
above the  $^2T$ ground state.  Unlike  the $^2T \to ^4A$  spin-flip excitation,
the excitation $^2T \to ^{2}H$, as  could have been perhaps argued from previous
theory   work~\cite{manini94,  obrien96,   Capone02,Capone04,Capone09,iwahara13}
takes place between spin doublets, and is thus of strictly orbital character ($L
=1  \to  L=2$). It  is  analogous  in this  respect  to  the $"d-d"$  "orbitons"
theorized~\cite{Ishihara97,  vanderBrink98} and  observed~\cite{Schlappa2012} at
much higher energy in transition  metal based Mott insulators. In Cs$_3$C$_{60}$
the large ion  size and the consequently small exchange  $J_{\rm H}$ lowers this
excitation down into  the range of nuclear vibrations  and dynamical Jahn-Teller
terms, with which it becomes entangled by hybridization.  Optically forbidden by
parity, this  orbital excitation  is low  enough in  energy to  become thermally
populated even below room temperature.

\subsection{Temperature dependent IR Spectrum}
Our first principles  approach yields quantitative predictions  for the infrared
spectra of  the Mott state, in  any of its  multiplet states.  Each state  has a
different IR  spectrum, directly  obtainable from  that, DFT-calculated,  of the
three adiabatic states.  First, $I_{^4\!A}=I_\alpha$, because the two states are
identical. Second,  from Eq.~\eqn{beta-rough} we obtain  that $2I_{^2\!H} \simeq
3I_\beta - I_\alpha\simeq 2I_\beta$.  The IR  spectrum of the ground state $^2T$
finally,  not calculable  exactly through  Eq.  \eqn{gamma-rough}  owing to  the
unknown interference  with $^2H$  in $I_\gamma$,  is reasonably  approximated by
$I_{^2T}\simeq I_\gamma$,
%%%%%%%%%%%%%%%%%%%%%%%%%%%%%%%%%%%%%%%%%%%%%%%%%%%%%%%%%%%%%%%%%%%%%%%%%%%%%%%%
\begin{figure}[htb!]
	\centering
	\includegraphics[width=0.95\linewidth]{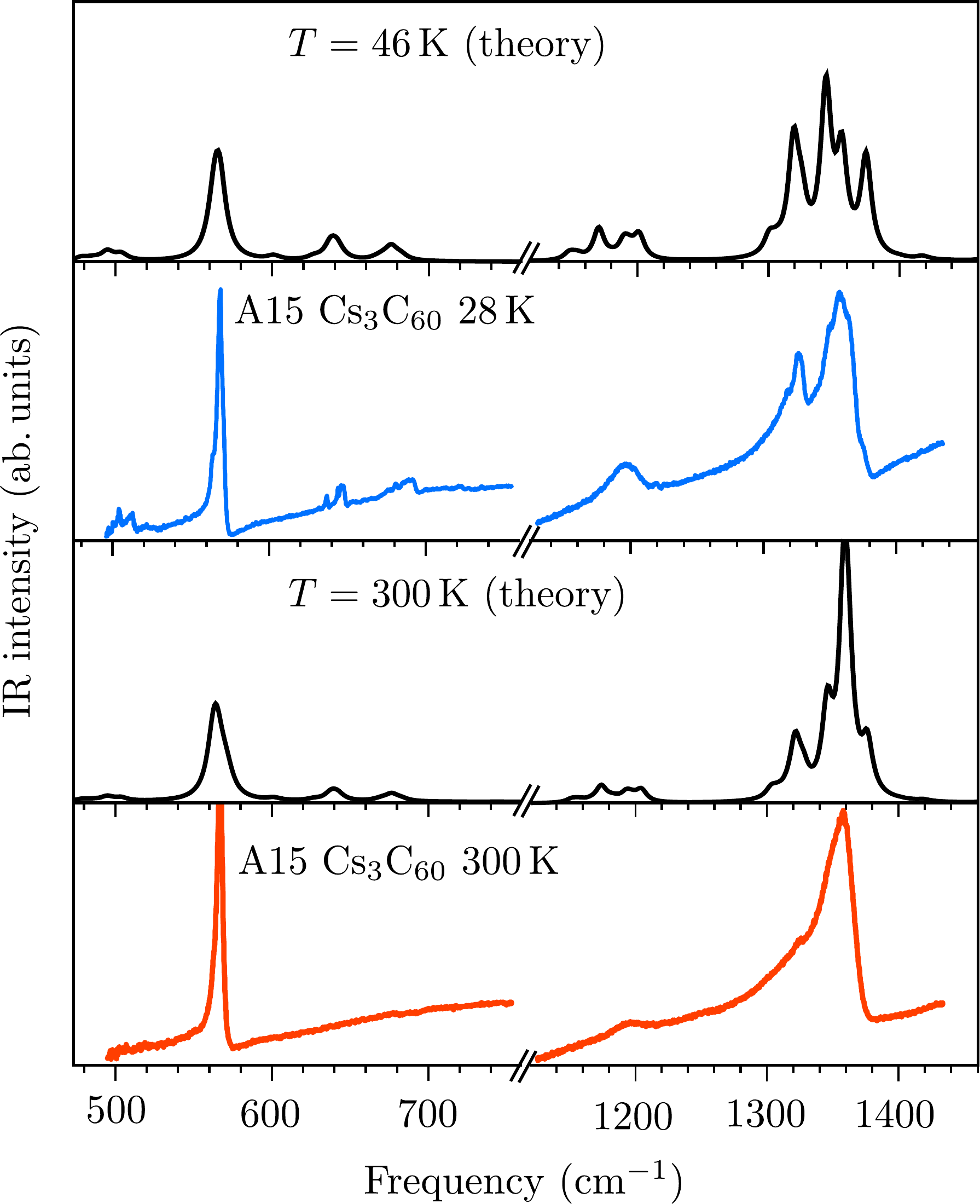}
	\caption{IR  spectra of  Mott  insulating  Cs$_{3}$C$_{60}$ (A15)  (from
Ref.~\cite{klupp12})  compared with  calculated ones  $I(\omega,T) =  \Sigma_{i=
^2T,  ^2H,^4A }  I_i(\omega) exp{-E_{i}/k_{B}T}  $ at  and above  $T_{\rm N}\sim
46\,K$, displayed  with an  arbitrary but reasonable  broadening width  $\sim 10
cm^-1$ The  thermal attenuation  of all $T_{\rm1u}(4)$  splittings and  of other
distortion-related peaks characteristic  of the $^2T$ ground state  is caused by
population of the orbital excitation $^2H$  rather than to simple washing out of
all distortions }
\end{figure}
%%%%%%%%%%%%%%%%%%%%%%%%%%%%%%%%%%%%%%%%%%%%%%%%%%%%%%%%%%%%%%%%%%%%%%%%%%%%%%
We finally compare in Fig.~3 these theoretical predictions with the experimental
IR spectra,  both at low  temperature and  in their temperature  evolution. Well
below room temperature the spectrum is dominated by the ground state $^2T$ whose
IR   is  similar   to  $I_{\gamma}   $,   sporting  heavy   traces  of   nuclear
distortions.~\cite{klupp12,kamaras13,  kamaras2014}  The  main  peak  near  1370
cm$^{-1}$, attributed to  T$_{\rm 1u}(4)$, is split in three  components, two of
which are  visible in  experiment (a third  one is actually  visible in  the fcc
structure).~\cite{klupp12,  kamaras13} Importantly,  new  peaks  near 640,  680,
1200, 1330  cm$^{-1}$ are prominent  in our calculation, in  good correspondence
with experiment at 28\,$K$ They are caused by B$_{\rm u}$ modes originating from
$H_u$ and some T$_{\rm 2u}$ ones, formerly  silent but now partly allowed by the
large size ground state nuclear distotion.

Around room  temperature instead we expect  the total thermal population  of the
tenfold degenerate  $^2H$ orbital excitation  to become comparable with  that of
the  sixfold-degenerate $^2T$  ground  state. Consequently  the  IR spectrum  is
increasingly  dominated  by  $I_{^2\!H}\simeq I_\beta$,  causing  the  resulting
intensity  demise  of splittings  and  new  peaks, that  are  distortion-related
features.  This temperature-induced demise, previously attributed to the thermal
washing out the classica adiabatic distortions (not unreasonably, but in reality
puzzlingly at the quantum level because of the much larger stabiliztion energies
signaled by the $\sim$\,800\,K $^2T\!\to\!^4\!A$ excitation across the spin gap) in
fact represents the signature of the  $^2T\to\, ^2H$ orbital excitation. Were it
not for the low-lying $^2H$  orbital excitation, the nuclear distortions, surely
still  strong at  room temperature,  would  not permit  the IR  extra peaks  and
splittings  to either  attenuate  or  disappear. The  thermal  frailty of  these
distortion-related  IR  features  represents  the  ``smoking  gun''  of  orbital
excitations in the molecular Mott insulator.

Before closing  we should  mention that additional  physical phenomena  to those
described so far will occur  below the N\'eel temperature, $T_{\rm N}\sim46\,K$.
In this  regime the symmetry-breaking  onset of antiferromagnetism  introduces a
Kugel-Khomskii    coupling    between    spin    and    orbital    degrees    of
freedom.~\cite{Capone09} Consequences will include the onset of a static orbital
distortion,  and a  corresponding magnon-orbiton  coupling. The  IR spectra  and
their description will also undergo weak  modifications, expected to be weak but
probably  not undetectable  experimentally. These  aspects are  left for  future
work.

\section{Conclusions}
In  summary,  we have  carried  out  a first  realization  of  a program,  where
low-lying multiplet states of molecular Mott insulating have been quantitatively
derived, and their signature characterized  spectroscopically for the first time
by  nanoscale  site  calculations  which  respect  the  full  orbital  symmetry,
generally violated by first-principles approaches.  The lowest excited state, an
important feature  of the  MMI so far  neglected, with  mixed electronic-nuclear
character but  purely orbital in character,  has been addressed and  shown to be
responsible for the experimental washing out of most distortion related features
observed  in IR  spectra  at  temperatures above  $\sim$  200--300\,K without  a
simultaneous diappearance of the overall  dynamic distortion. A similar fate and
role  of  an orbital  excitation,  which  is low-lying  in  A15-Cs$_{3}$C$_{60}$
because of the nanoscale as opposed to atomic size of the insulating site, could
be pursued in other MMIs in the future.  On the given system, on the other hand,
other properties  including EPR spectra would  very likely be influenced  by the
dynamic orbital physics outlined here.

%%%%%%%%%%%%%%%%%%%%%%%%%%%%%%%%%%%%%%%%%%%%%%%%%%%%%%%%%%%%%%%%%%%%%%%%%%%%%%%%%
%%%%%%%%%%%%%%%%%%%%%%%%%%%%%%%%%%%%%%%%%%%%%%%%%%%%%%%%%%%%%%%%%%%%%%%%%%%%%%%%%
\begin{acknowledgments}  
This  work was  supported by  the European  Union FP7-NMP-2011-EU-Japan  Project
LEMSUPER, in part by  ERC Advanced Grant MODPHYSFRICT and by  a CINECA HPC award
2013.  We are grateful to K.  Prassides for correcting an error in our initially
assumed crystal structure, 
and to G. Klupp, K. Kamaras, D. Arcon for information and discussions.
\end{acknowledgments}


\begin{thebibliography}{37}%
\makeatletter
\providecommand \@ifxundefined [1]{%
 \@ifx{#1\undefined}
}%
\providecommand \@ifnum [1]{%
 \ifnum #1\expandafter \@firstoftwo
 \else \expandafter \@secondoftwo
 \fi
}%
\providecommand \@ifx [1]{%
 \ifx #1\expandafter \@firstoftwo
 \else \expandafter \@secondoftwo
 \fi
}%
\providecommand \natexlab [1]{#1}%
\providecommand \enquote  [1]{``#1''}%
\providecommand \bibnamefont  [1]{#1}%
\providecommand \bibfnamefont [1]{#1}%
\providecommand \citenamefont [1]{#1}%
\providecommand \href@noop [0]{\@secondoftwo}%
\providecommand \href [0]{\begingroup \@sanitize@url \@href}%
\providecommand \@href[1]{\@@startlink{#1}\@@href}%
\providecommand \@@href[1]{\endgroup#1\@@endlink}%
\providecommand \@sanitize@url [0]{\catcode `\\12\catcode `\$12\catcode
  `\&12\catcode `\#12\catcode `\^12\catcode `\_12\catcode `\%12\relax}%
\providecommand \@@startlink[1]{}%
\providecommand \@@endlink[0]{}%
\providecommand \url  [0]{\begingroup\@sanitize@url \@url }%
\providecommand \@url [1]{\endgroup\@href {#1}{\urlprefix }}%
\providecommand \urlprefix  [0]{URL }%
\providecommand \Eprint [0]{\href }%
\providecommand \doibase [0]{http://dx.doi.org/}%
\providecommand \selectlanguage [0]{\@gobble}%
\providecommand \bibinfo  [0]{\@secondoftwo}%
\providecommand \bibfield  [0]{\@secondoftwo}%
\providecommand \translation [1]{[#1]}%
\providecommand \BibitemOpen [0]{}%
\providecommand \bibitemStop [0]{}%
\providecommand \bibitemNoStop [0]{.\EOS\space}%
\providecommand \EOS [0]{\spacefactor3000\relax}%
\providecommand \BibitemShut  [1]{\csname bibitem#1\endcsname}%
\let\auto@bib@innerbib\@empty
%</preamble>
\bibitem [{\citenamefont {Goodenough}(1971)}]{Goodenough1971}%
  \BibitemOpen
  \bibfield  {author} {\bibinfo {author} {\bibfnamefont {J.~B.}\ \bibnamefont
  {Goodenough}},\ }\href {\doibase 10.1016/0022-4596(71)90091-0} {\bibfield
  {journal} {\bibinfo  {journal} {J. Solid State Chem.}\ }\textbf {\bibinfo
  {volume} {3}},\ \bibinfo {pages} {490} (\bibinfo {year} {1971})}\BibitemShut
  {NoStop}%
\bibitem [{\citenamefont {Kugel'}\ and\ \citenamefont
  {Khomski}(1982)}]{Kugel-Khomskii82}%
  \BibitemOpen
  \bibfield  {author} {\bibinfo {author} {\bibfnamefont {K.~I.}\ \bibnamefont
  {Kugel'}}\ and\ \bibinfo {author} {\bibfnamefont {D.~I.}\ \bibnamefont
  {Khomski}},\ }\href {\doibase 10.1070/PU1982v025n04ABEH004537} {\bibfield
  {journal} {\bibinfo  {journal} {Sov.~Phys.~Usp.}\ }\textbf {\bibinfo {volume}
  {25}},\ \bibinfo {pages} {231} (\bibinfo {year} {1982})}\BibitemShut
  {NoStop}%
\bibitem [{\citenamefont {Schlappa}\ \emph {et~al.}(2012)\citenamefont
  {Schlappa}, \citenamefont {Wohlfeld}, \citenamefont {Zhou}, \citenamefont
  {Mourigal}, \citenamefont {Haverkort}, \citenamefont {Strocov}, \citenamefont
  {Hozoi}, \citenamefont {Monney}, \citenamefont {Nishimoto}, \citenamefont
  {Singh}, \citenamefont {Revcolevschi}, \citenamefont {Caux}, \citenamefont
  {Patthey}, \citenamefont {R{\o}nnow}, \citenamefont {van~den Brink},\ and\
  \citenamefont {Schmitt}}]{Schlappa2012}%
  \BibitemOpen
  \bibfield  {author} {\bibinfo {author} {\bibfnamefont {J.}~\bibnamefont
  {Schlappa}}, \bibinfo {author} {\bibfnamefont {K.}~\bibnamefont {Wohlfeld}},
  \bibinfo {author} {\bibfnamefont {K.~J.}\ \bibnamefont {Zhou}}, \bibinfo
  {author} {\bibfnamefont {M.}~\bibnamefont {Mourigal}}, \bibinfo {author}
  {\bibfnamefont {M.~W.}\ \bibnamefont {Haverkort}}, \bibinfo {author}
  {\bibfnamefont {V.~N.}\ \bibnamefont {Strocov}}, \bibinfo {author}
  {\bibfnamefont {L.}~\bibnamefont {Hozoi}}, \bibinfo {author} {\bibfnamefont
  {C.}~\bibnamefont {Monney}}, \bibinfo {author} {\bibfnamefont
  {S.}~\bibnamefont {Nishimoto}}, \bibinfo {author} {\bibfnamefont
  {S.}~\bibnamefont {Singh}}, \bibinfo {author} {\bibfnamefont
  {A.}~\bibnamefont {Revcolevschi}}, \bibinfo {author} {\bibfnamefont {J.-S.}\
  \bibnamefont {Caux}}, \bibinfo {author} {\bibfnamefont {L.}~\bibnamefont
  {Patthey}}, \bibinfo {author} {\bibfnamefont {H.~M.}\ \bibnamefont
  {R{\o}nnow}}, \bibinfo {author} {\bibfnamefont {J.}~\bibnamefont {van~den
  Brink}}, \ and\ \bibinfo {author} {\bibfnamefont {T.}~\bibnamefont
  {Schmitt}},\ }\href {\doibase 10.1038/nature10974} {\bibfield  {journal}
  {\bibinfo  {journal} {Nature}\ }\textbf {\bibinfo {volume} {485}},\ \bibinfo
  {pages} {82} (\bibinfo {year} {2012})}\BibitemShut {NoStop}%
\bibitem [{\citenamefont {Castellani}\ \emph {et~al.}(1978)\citenamefont
  {Castellani}, \citenamefont {Natoli},\ and\ \citenamefont
  {Ranninger}}]{Castellani1978}%
  \BibitemOpen
  \bibfield  {author} {\bibinfo {author} {\bibfnamefont {C.}~\bibnamefont
  {Castellani}}, \bibinfo {author} {\bibfnamefont {C.~R.}\ \bibnamefont
  {Natoli}}, \ and\ \bibinfo {author} {\bibfnamefont {J.}~\bibnamefont
  {Ranninger}},\ }\href {\doibase 10.1103/PhysRevB.18.4945} {\bibfield
  {journal} {\bibinfo  {journal} {Phys. Rev. B}\ }\textbf {\bibinfo {volume}
  {18}},\ \bibinfo {pages} {4945} (\bibinfo {year} {1978})}\BibitemShut
  {NoStop}%
\bibitem [{\citenamefont {Ishihara}\ \emph
  {et~al.}(1997{\natexlab{a}})\citenamefont {Ishihara}, \citenamefont
  {Yamanaka},\ and\ \citenamefont {Nagaosa}}]{Ishihara1997}%
  \BibitemOpen
  \bibfield  {author} {\bibinfo {author} {\bibfnamefont {S.}~\bibnamefont
  {Ishihara}}, \bibinfo {author} {\bibfnamefont {M.}~\bibnamefont {Yamanaka}},
  \ and\ \bibinfo {author} {\bibfnamefont {N.}~\bibnamefont {Nagaosa}},\ }\href
  {\doibase 10.1103/PhysRevB.56.686} {\bibfield  {journal} {\bibinfo  {journal}
  {Phys. Rev. B}\ }\textbf {\bibinfo {volume} {56}},\ \bibinfo {pages} {686}
  (\bibinfo {year} {1997}{\natexlab{a}})}\BibitemShut {NoStop}%
\bibitem [{\citenamefont {van~den Brink}\ and\ \citenamefont
  {Khomskii}(2001)}]{VandenBrink2001}%
  \BibitemOpen
  \bibfield  {author} {\bibinfo {author} {\bibfnamefont {J.}~\bibnamefont
  {van~den Brink}}\ and\ \bibinfo {author} {\bibfnamefont {D.}~\bibnamefont
  {Khomskii}},\ }\href {\doibase 10.1103/PhysRevB.63.140416} {\bibfield
  {journal} {\bibinfo  {journal} {Phys. Rev. B}\ }\textbf {\bibinfo {volume}
  {63}},\ \bibinfo {pages} {140416} (\bibinfo {year} {2001})}\BibitemShut
  {NoStop}%
\bibitem [{\citenamefont {Khaliullin}\ and\ \citenamefont
  {Maekawa}(2000)}]{Khaliullin2000}%
  \BibitemOpen
  \bibfield  {author} {\bibinfo {author} {\bibfnamefont {G.}~\bibnamefont
  {Khaliullin}}\ and\ \bibinfo {author} {\bibfnamefont {S.}~\bibnamefont
  {Maekawa}},\ }\href {\doibase 10.1103/PhysRevLett.85.3950} {\bibfield
  {journal} {\bibinfo  {journal} {Phys. Rev. Lett.}\ }\textbf {\bibinfo
  {volume} {85}},\ \bibinfo {pages} {3950} (\bibinfo {year}
  {2000})}\BibitemShut {NoStop}%
\bibitem [{\citenamefont {Ganin}\ \emph {et~al.}(2008)\citenamefont {Ganin},
  \citenamefont {Takabayashi}, \citenamefont {Khimyak}, \citenamefont
  {Margadonna}, \citenamefont {Tamai}, \citenamefont {Rosseinsky},\ and\
  \citenamefont {Prassides}}]{ganin08}%
  \BibitemOpen
  \bibfield  {author} {\bibinfo {author} {\bibfnamefont {A.~Y.}\ \bibnamefont
  {Ganin}}, \bibinfo {author} {\bibfnamefont {Y.}~\bibnamefont {Takabayashi}},
  \bibinfo {author} {\bibfnamefont {Y.~Z.}\ \bibnamefont {Khimyak}}, \bibinfo
  {author} {\bibfnamefont {S.}~\bibnamefont {Margadonna}}, \bibinfo {author}
  {\bibfnamefont {A.}~\bibnamefont {Tamai}}, \bibinfo {author} {\bibfnamefont
  {M.~J.}\ \bibnamefont {Rosseinsky}}, \ and\ \bibinfo {author} {\bibfnamefont
  {K.}~\bibnamefont {Prassides}},\ }\href {\doibase 10.1038/nmat2179}
  {\bibfield  {journal} {\bibinfo  {journal} {nature materials}\ }\textbf
  {\bibinfo {volume} {7}},\ \bibinfo {pages} {367} (\bibinfo {year}
  {2008})}\BibitemShut {NoStop}%
\bibitem [{\citenamefont {Takabayashi}\ \emph {et~al.}(2009)\citenamefont
  {Takabayashi}, \citenamefont {Ganin}, \citenamefont {Jegli\v{c}},
  \citenamefont {Ar\v{c}on}, \citenamefont {Takano}, \citenamefont {Iwasa},
  \citenamefont {nad}, \citenamefont {Takata}, \citenamefont {Takeshita},
  \citenamefont {Prassides},\ and\ \citenamefont {Rosseinsky}}]{Takabayashi09}%
  \BibitemOpen
  \bibfield  {author} {\bibinfo {author} {\bibfnamefont {Y.}~\bibnamefont
  {Takabayashi}}, \bibinfo {author} {\bibfnamefont {A.~Y.}\ \bibnamefont
  {Ganin}}, \bibinfo {author} {\bibfnamefont {P.}~\bibnamefont {Jegli\v{c}}},
  \bibinfo {author} {\bibfnamefont {D.}~\bibnamefont {Ar\v{c}on}}, \bibinfo
  {author} {\bibfnamefont {T.}~\bibnamefont {Takano}}, \bibinfo {author}
  {\bibfnamefont {Y.}~\bibnamefont {Iwasa}}, \bibinfo {author} {\bibfnamefont
  {Y.~O.}\ \bibnamefont {nad}}, \bibinfo {author} {\bibfnamefont
  {M.}~\bibnamefont {Takata}}, \bibinfo {author} {\bibfnamefont
  {N.}~\bibnamefont {Takeshita}}, \bibinfo {author} {\bibfnamefont
  {K.}~\bibnamefont {Prassides}}, \ and\ \bibinfo {author} {\bibfnamefont
  {M.~J.}\ \bibnamefont {Rosseinsky}},\ }\href {\doibase
  10.1126/science.1169163} {\bibfield  {journal} {\bibinfo  {journal}
  {Science}\ }\textbf {\bibinfo {volume} {323}},\ \bibinfo {pages} {1585}
  (\bibinfo {year} {2009})}\BibitemShut {NoStop}%
\bibitem [{\citenamefont {Ihara}\ \emph {et~al.}(2010)\citenamefont {Ihara},
  \citenamefont {Alloul}, \citenamefont {Wzietek}, \citenamefont {Pontiroli},
  \citenamefont {Mazzani},\ and\ \citenamefont {Ricc\`o}}]{Ihara10}%
  \BibitemOpen
  \bibfield  {author} {\bibinfo {author} {\bibfnamefont {Y.}~\bibnamefont
  {Ihara}}, \bibinfo {author} {\bibfnamefont {H.}~\bibnamefont {Alloul}},
  \bibinfo {author} {\bibfnamefont {P.}~\bibnamefont {Wzietek}}, \bibinfo
  {author} {\bibfnamefont {D.}~\bibnamefont {Pontiroli}}, \bibinfo {author}
  {\bibfnamefont {M.}~\bibnamefont {Mazzani}}, \ and\ \bibinfo {author}
  {\bibfnamefont {M.}~\bibnamefont {Ricc\`o}},\ }\href {\doibase
  10.1103/PhysRevLett.104.256402} {\bibfield  {journal} {\bibinfo  {journal}
  {Phys. Rev. Lett.}\ }\textbf {\bibinfo {volume} {104}},\ \bibinfo {pages}
  {256402} (\bibinfo {year} {2010})}\BibitemShut {NoStop}%
\bibitem [{\citenamefont {Ganin}\ \emph {et~al.}(2010)\citenamefont {Ganin},
  \citenamefont {Takabayashi}, \citenamefont {Jegli\v{c}}, \citenamefont
  {Ar\v{c}on}, \citenamefont {Poto\v{c}nik}, \citenamefont {Baker},
  \citenamefont {Ohishi}, \citenamefont {McDonald}, \citenamefont {Tzirakis},
  \citenamefont {Mc{L}ennan}, \citenamefont {Darling}, \citenamefont {Takata},
  \citenamefont {Rosseinsky},\ and\ \citenamefont {Prassides}}]{Ganin10}%
  \BibitemOpen
  \bibfield  {author} {\bibinfo {author} {\bibfnamefont {A.~Y.}\ \bibnamefont
  {Ganin}}, \bibinfo {author} {\bibfnamefont {Y.}~\bibnamefont {Takabayashi}},
  \bibinfo {author} {\bibfnamefont {P.}~\bibnamefont {Jegli\v{c}}}, \bibinfo
  {author} {\bibfnamefont {D.}~\bibnamefont {Ar\v{c}on}}, \bibinfo {author}
  {\bibfnamefont {A.}~\bibnamefont {Poto\v{c}nik}}, \bibinfo {author}
  {\bibfnamefont {P.~J.}\ \bibnamefont {Baker}}, \bibinfo {author}
  {\bibfnamefont {Y.}~\bibnamefont {Ohishi}}, \bibinfo {author} {\bibfnamefont
  {M.~T.}\ \bibnamefont {McDonald}}, \bibinfo {author} {\bibfnamefont {M.~D.}\
  \bibnamefont {Tzirakis}}, \bibinfo {author} {\bibfnamefont {A.}~\bibnamefont
  {Mc{L}ennan}}, \bibinfo {author} {\bibfnamefont {G.~R.}\ \bibnamefont
  {Darling}}, \bibinfo {author} {\bibfnamefont {M.}~\bibnamefont {Takata}},
  \bibinfo {author} {\bibfnamefont {M.~J.}\ \bibnamefont {Rosseinsky}}, \ and\
  \bibinfo {author} {\bibfnamefont {K.}~\bibnamefont {Prassides}},\ }\href
  {\doibase 10.1038/nature09120} {\bibfield  {journal} {\bibinfo  {journal}
  {Nature}\ }\textbf {\bibinfo {volume} {466}},\ \bibinfo {pages} {221}
  (\bibinfo {year} {2010})}\BibitemShut {NoStop}%
\bibitem [{\citenamefont {Kamar{\'{a}}s}\ and\ \citenamefont
  {Klupp}(2014)}]{kamaras2014}%
  \BibitemOpen
  \bibfield  {author} {\bibinfo {author} {\bibfnamefont {K.}~\bibnamefont
  {Kamar{\'{a}}s}}\ and\ \bibinfo {author} {\bibfnamefont {G.}~\bibnamefont
  {Klupp}},\ }\href {\doibase 10.1039/c4dt00206g} {\bibfield  {journal}
  {\bibinfo  {journal} {Dalt. Trans.}\ }\textbf {\bibinfo {volume} {43}},\
  \bibinfo {pages} {7366} (\bibinfo {year} {2014})}\BibitemShut {NoStop}%
\bibitem [{\citenamefont {Zadik}\ \emph {et~al.}(2015)\citenamefont {Zadik},
  \citenamefont {Takabayashi}, \citenamefont {Klupp}, \citenamefont {Colman},
  \citenamefont {Ganin}, \citenamefont {Potocnik}, \citenamefont {Jeglic},
  \citenamefont {Arcon}, \citenamefont {Matus}, \citenamefont {Kamaras},
  \citenamefont {Kasahara}, \citenamefont {Iwasa}, \citenamefont {Fitch},
  \citenamefont {Ohishi}, \citenamefont {Garbarino}, \citenamefont {Kato},
  \citenamefont {Rosseinsky},\ and\ \citenamefont {Prassides}}]{Zadik2015}%
  \BibitemOpen
  \bibfield  {author} {\bibinfo {author} {\bibfnamefont {R.~H.}\ \bibnamefont
  {Zadik}}, \bibinfo {author} {\bibfnamefont {Y.}~\bibnamefont {Takabayashi}},
  \bibinfo {author} {\bibfnamefont {G.}~\bibnamefont {Klupp}}, \bibinfo
  {author} {\bibfnamefont {R.~H.}\ \bibnamefont {Colman}}, \bibinfo {author}
  {\bibfnamefont {A.~Y.}\ \bibnamefont {Ganin}}, \bibinfo {author}
  {\bibfnamefont {A.}~\bibnamefont {Potocnik}}, \bibinfo {author}
  {\bibfnamefont {P.}~\bibnamefont {Jeglic}}, \bibinfo {author} {\bibfnamefont
  {D.}~\bibnamefont {Arcon}}, \bibinfo {author} {\bibfnamefont
  {P.}~\bibnamefont {Matus}}, \bibinfo {author} {\bibfnamefont
  {K.}~\bibnamefont {Kamaras}}, \bibinfo {author} {\bibfnamefont
  {Y.}~\bibnamefont {Kasahara}}, \bibinfo {author} {\bibfnamefont
  {Y.}~\bibnamefont {Iwasa}}, \bibinfo {author} {\bibfnamefont {A.~N.}\
  \bibnamefont {Fitch}}, \bibinfo {author} {\bibfnamefont {Y.}~\bibnamefont
  {Ohishi}}, \bibinfo {author} {\bibfnamefont {G.}~\bibnamefont {Garbarino}},
  \bibinfo {author} {\bibfnamefont {K.}~\bibnamefont {Kato}}, \bibinfo {author}
  {\bibfnamefont {M.~J.}\ \bibnamefont {Rosseinsky}}, \ and\ \bibinfo {author}
  {\bibfnamefont {K.}~\bibnamefont {Prassides}},\ }\href {\doibase
  10.1126/sciadv.1500059} {\bibfield  {journal} {\bibinfo  {journal} {Sci.
  Adv.}\ }\textbf {\bibinfo {volume} {1}},\ \bibinfo {pages} {e1500059}
  (\bibinfo {year} {2015})}\BibitemShut {NoStop}%
\bibitem [{\citenamefont {Baldassarre}\ \emph {et~al.}(2015)\citenamefont
  {Baldassarre}, \citenamefont {Perucchi}, \citenamefont {Mitrano},
  \citenamefont {Nicoletti}, \citenamefont {Marini}, \citenamefont {Pontiroli},
  \citenamefont {Mazzani}, \citenamefont {Aramini}, \citenamefont
  {Ricc{\'{o}}}, \citenamefont {Giovannetti}, \citenamefont {Capone},\ and\
  \citenamefont {Lupi}}]{Baldassarre2015}%
  \BibitemOpen
  \bibfield  {author} {\bibinfo {author} {\bibfnamefont {L.}~\bibnamefont
  {Baldassarre}}, \bibinfo {author} {\bibfnamefont {A.}~\bibnamefont
  {Perucchi}}, \bibinfo {author} {\bibfnamefont {M.}~\bibnamefont {Mitrano}},
  \bibinfo {author} {\bibfnamefont {D.}~\bibnamefont {Nicoletti}}, \bibinfo
  {author} {\bibfnamefont {C.}~\bibnamefont {Marini}}, \bibinfo {author}
  {\bibfnamefont {D.}~\bibnamefont {Pontiroli}}, \bibinfo {author}
  {\bibfnamefont {M.}~\bibnamefont {Mazzani}}, \bibinfo {author} {\bibfnamefont
  {M.}~\bibnamefont {Aramini}}, \bibinfo {author} {\bibfnamefont
  {M.}~\bibnamefont {Ricc{\'{o}}}}, \bibinfo {author} {\bibfnamefont
  {G.}~\bibnamefont {Giovannetti}}, \bibinfo {author} {\bibfnamefont
  {M.}~\bibnamefont {Capone}}, \ and\ \bibinfo {author} {\bibfnamefont
  {S.}~\bibnamefont {Lupi}},\ }\href {\doibase 10.1038/srep15240} {\bibfield
  {journal} {\bibinfo  {journal} {Sci. Rep.}\ }\textbf {\bibinfo {volume}
  {5}},\ \bibinfo {pages} {15240} (\bibinfo {year} {2015})}\BibitemShut
  {NoStop}%
\bibitem [{\citenamefont {Gunnarsson}(1997)}]{Gunnarsson97}%
  \BibitemOpen
  \bibfield  {author} {\bibinfo {author} {\bibfnamefont {O.}~\bibnamefont
  {Gunnarsson}},\ }\href {\doibase 10.1103/RevModPhys.69.575} {\bibfield
  {journal} {\bibinfo  {journal} {Rev. Mod. Phys.}\ }\textbf {\bibinfo {volume}
  {69}},\ \bibinfo {pages} {575} (\bibinfo {year} {1997})}\BibitemShut
  {NoStop}%
\bibitem [{\citenamefont {Jegli\v{c}}\ \emph {et~al.}(2009)\citenamefont
  {Jegli\v{c}}, \citenamefont {Ar\v{c}on}, \citenamefont {Poto\u{c}nik},
  \citenamefont {Ganin}, \citenamefont {Takabayashi}, \citenamefont
  {Rosseinsky},\ and\ \citenamefont {Prassides}}]{Jeglic09}%
  \BibitemOpen
  \bibfield  {author} {\bibinfo {author} {\bibfnamefont {P.}~\bibnamefont
  {Jegli\v{c}}}, \bibinfo {author} {\bibfnamefont {D.}~\bibnamefont
  {Ar\v{c}on}}, \bibinfo {author} {\bibfnamefont {A.}~\bibnamefont
  {Poto\u{c}nik}}, \bibinfo {author} {\bibfnamefont {A.~Y.}\ \bibnamefont
  {Ganin}}, \bibinfo {author} {\bibfnamefont {Y.}~\bibnamefont {Takabayashi}},
  \bibinfo {author} {\bibfnamefont {M.~J.}\ \bibnamefont {Rosseinsky}}, \ and\
  \bibinfo {author} {\bibfnamefont {K.}~\bibnamefont {Prassides}},\ }\href
  {\doibase 10.1103/PhysRevB.80.195424} {\bibfield  {journal} {\bibinfo
  {journal} {Phys. Rev. B}\ }\textbf {\bibinfo {volume} {80}},\ \bibinfo
  {pages} {195424} (\bibinfo {year} {2009})}\BibitemShut {NoStop}%
\bibitem [{\citenamefont {Wzietek}\ \emph {et~al.}(2014)\citenamefont
  {Wzietek}, \citenamefont {Mito}, \citenamefont {Alloul}, \citenamefont
  {Pontiroli}, \citenamefont {Aramini},\ and\ \citenamefont
  {Ricc\`o}}]{Wzietek14}%
  \BibitemOpen
  \bibfield  {author} {\bibinfo {author} {\bibfnamefont {P.}~\bibnamefont
  {Wzietek}}, \bibinfo {author} {\bibfnamefont {T.}~\bibnamefont {Mito}},
  \bibinfo {author} {\bibfnamefont {H.}~\bibnamefont {Alloul}}, \bibinfo
  {author} {\bibfnamefont {D.}~\bibnamefont {Pontiroli}}, \bibinfo {author}
  {\bibfnamefont {M.}~\bibnamefont {Aramini}}, \ and\ \bibinfo {author}
  {\bibfnamefont {M.}~\bibnamefont {Ricc\`o}},\ }\href {\doibase
  10.1103/PhysRevLett.112.066401} {\bibfield  {journal} {\bibinfo  {journal}
  {Phys. Rev. Lett.}\ }\textbf {\bibinfo {volume} {112}},\ \bibinfo {pages}
  {066401} (\bibinfo {year} {2014})}\BibitemShut {NoStop}%
\bibitem [{\citenamefont {Chibotaru}(2005)}]{chibotaru05}%
  \BibitemOpen
  \bibfield  {author} {\bibinfo {author} {\bibfnamefont {L.~F.}\ \bibnamefont
  {Chibotaru}},\ }\href {\doibase 10.1103/PhysRevLett.94.186405} {\bibfield
  {journal} {\bibinfo  {journal} {Phys. Rev. Lett.}\ }\textbf {\bibinfo
  {volume} {94}},\ \bibinfo {pages} {186405} (\bibinfo {year}
  {2005})}\BibitemShut {NoStop}%
\bibitem [{\citenamefont {Iwahara}\ and\ \citenamefont
  {Chibotaru}(2013)}]{iwahara13}%
  \BibitemOpen
  \bibfield  {author} {\bibinfo {author} {\bibfnamefont {N.}~\bibnamefont
  {Iwahara}}\ and\ \bibinfo {author} {\bibfnamefont {L.~F.}\ \bibnamefont
  {Chibotaru}},\ }\href {\doibase 10.1103/PhysRevLett.111.056401} {\bibfield
  {journal} {\bibinfo  {journal} {Phys. Rev. Lett.}\ }\textbf {\bibinfo
  {volume} {111}},\ \bibinfo {pages} {056401} (\bibinfo {year}
  {2013})}\BibitemShut {NoStop}%
\bibitem [{\citenamefont {Iwahara}\ and\ \citenamefont
  {Chibotaru}(2015)}]{iwahara15}%
  \BibitemOpen
  \bibfield  {author} {\bibinfo {author} {\bibfnamefont {N.}~\bibnamefont
  {Iwahara}}\ and\ \bibinfo {author} {\bibfnamefont {L.~F.}\ \bibnamefont
  {Chibotaru}},\ }\href {\doibase 10.1103/PhysRevB.91.035109} {\bibfield
  {journal} {\bibinfo  {journal} {Phys. Rev. B}\ }\textbf {\bibinfo {volume}
  {91}},\ \bibinfo {pages} {035109} (\bibinfo {year} {2015})}\BibitemShut
  {NoStop}%
\bibitem [{\citenamefont {Georges}\ \emph {et~al.}(1996)\citenamefont
  {Georges}, \citenamefont {Kotliar}, \citenamefont {Krauth},\ and\
  \citenamefont {Rozenberg}}]{Georges1996}%
  \BibitemOpen
  \bibfield  {author} {\bibinfo {author} {\bibfnamefont {A.}~\bibnamefont
  {Georges}}, \bibinfo {author} {\bibfnamefont {G.}~\bibnamefont {Kotliar}},
  \bibinfo {author} {\bibfnamefont {W.}~\bibnamefont {Krauth}}, \ and\ \bibinfo
  {author} {\bibfnamefont {M.}~\bibnamefont {Rozenberg}},\ }\href {\doibase
  http://dx.doi.org/10.1103/RevModPhys.68.13} {\bibfield  {journal} {\bibinfo
  {journal} {Rev. Mod. Phys.}\ }\textbf {\bibinfo {volume} {68}},\ \bibinfo
  {pages} {13} (\bibinfo {year} {1996})}\BibitemShut {NoStop}%
\bibitem [{\citenamefont {Gunnarsson}\ \emph {et~al.}(1996)\citenamefont
  {Gunnarsson}, \citenamefont {Koch},\ and\ \citenamefont
  {Martin}}]{Gunnarssson96}%
  \BibitemOpen
  \bibfield  {author} {\bibinfo {author} {\bibfnamefont {O.}~\bibnamefont
  {Gunnarsson}}, \bibinfo {author} {\bibfnamefont {E.}~\bibnamefont {Koch}}, \
  and\ \bibinfo {author} {\bibfnamefont {R.~M.}\ \bibnamefont {Martin}},\
  }\href {\doibase 10.1103/PhysRevB.54.R11026} {\bibfield  {journal} {\bibinfo
  {journal} {Phys. Rev. B}\ }\textbf {\bibinfo {volume} {54}},\ \bibinfo
  {pages} {R11026} (\bibinfo {year} {1996})}\BibitemShut {NoStop}%
\bibitem [{\citenamefont {Capone}\ \emph {et~al.}(2002)\citenamefont {Capone},
  \citenamefont {Fabrizio}, \citenamefont {Castellani},\ and\ \citenamefont
  {Tosatti}}]{Capone02}%
  \BibitemOpen
  \bibfield  {author} {\bibinfo {author} {\bibfnamefont {M.}~\bibnamefont
  {Capone}}, \bibinfo {author} {\bibfnamefont {M.}~\bibnamefont {Fabrizio}},
  \bibinfo {author} {\bibfnamefont {C.}~\bibnamefont {Castellani}}, \ and\
  \bibinfo {author} {\bibfnamefont {E.}~\bibnamefont {Tosatti}},\ }\href
  {\doibase 10.1126/science.1071122} {\bibfield  {journal} {\bibinfo  {journal}
  {Science}\ }\textbf {\bibinfo {volume} {296}},\ \bibinfo {pages} {2364}
  (\bibinfo {year} {2002})}\BibitemShut {NoStop}%
\bibitem [{\citenamefont {M.Capone}\ \emph {et~al.}(2009)\citenamefont
  {M.Capone}, \citenamefont {M.Fabrizio}, \citenamefont {C.Castellani},\ and\
  \citenamefont {E.Tosatti}}]{Capone09}%
  \BibitemOpen
  \bibfield  {author} {\bibinfo {author} {\bibnamefont {M.Capone}}, \bibinfo
  {author} {\bibnamefont {M.Fabrizio}}, \bibinfo {author} {\bibnamefont
  {C.Castellani}}, \ and\ \bibinfo {author} {\bibnamefont {E.Tosatti}},\ }\href
  {\doibase 10.1103/RevModPhys.81.943} {\bibfield  {journal} {\bibinfo
  {journal} {Rev.~Mod.~Phys.}\ }\textbf {\bibinfo {volume} {81}},\ \bibinfo
  {pages} {943} (\bibinfo {year} {2009})}\BibitemShut {NoStop}%
\bibitem [{\citenamefont {Nomura}\ \emph {et~al.}(2016)\citenamefont {Nomura},
  \citenamefont {Sakai}, \citenamefont {Capone},\ and\ \citenamefont
  {Arita}}]{Nomura16}%
  \BibitemOpen
  \bibfield  {author} {\bibinfo {author} {\bibfnamefont {Y.}~\bibnamefont
  {Nomura}}, \bibinfo {author} {\bibfnamefont {S.}~\bibnamefont {Sakai}},
  \bibinfo {author} {\bibfnamefont {M.}~\bibnamefont {Capone}}, \ and\ \bibinfo
  {author} {\bibfnamefont {R.}~\bibnamefont {Arita}},\ }\href {\doibase
  10.1088/0953-8984/28/15/153001} {\bibfield  {journal} {\bibinfo  {journal}
  {J. Phys. Condens. Matter}\ }\textbf {\bibinfo {volume} {28}},\ \bibinfo
  {pages} {153001} (\bibinfo {year} {2016})}\BibitemShut {NoStop}%
\bibitem [{\citenamefont {Negri}\ \emph {et~al.}(1992)\citenamefont {Negri},
  \citenamefont {Orlandi},\ and\ \citenamefont {Zerbetto}}]{Negri1992}%
  \BibitemOpen
  \bibfield  {author} {\bibinfo {author} {\bibfnamefont {F.}~\bibnamefont
  {Negri}}, \bibinfo {author} {\bibfnamefont {G.}~\bibnamefont {Orlandi}}, \
  and\ \bibinfo {author} {\bibfnamefont {F.}~\bibnamefont {Zerbetto}},\ }\href
  {\doibase 10.1016/0009-2614(92)85239-7} {\bibfield  {journal} {\bibinfo
  {journal} {Chem. Phys. Lett.}\ }\textbf {\bibinfo {volume} {189}},\ \bibinfo
  {pages} {495} (\bibinfo {year} {1992})}\BibitemShut {NoStop}%
\bibitem [{\citenamefont {Auerbach}\ \emph {et~al.}(1994)\citenamefont
  {Auerbach}, \citenamefont {Manini},\ and\ \citenamefont
  {Tosatti}}]{Auerbach1994}%
  \BibitemOpen
  \bibfield  {author} {\bibinfo {author} {\bibfnamefont {A.}~\bibnamefont
  {Auerbach}}, \bibinfo {author} {\bibfnamefont {N.}~\bibnamefont {Manini}}, \
  and\ \bibinfo {author} {\bibfnamefont {E.}~\bibnamefont {Tosatti}},\ }\href
  {\doibase 10.1103/PhysRevB.49.12998} {\bibfield  {journal} {\bibinfo
  {journal} {Phys. Rev. B}\ }\textbf {\bibinfo {volume} {49}},\ \bibinfo
  {pages} {12998} (\bibinfo {year} {1994})}\BibitemShut {NoStop}%
\bibitem [{\citenamefont {O'Brien}(1996)}]{obrien96}%
  \BibitemOpen
  \bibfield  {author} {\bibinfo {author} {\bibfnamefont {M.~C.~M.}\
  \bibnamefont {O'Brien}},\ }\href {\doibase 10.1103/PhysRevB.53.3775}
  {\bibfield  {journal} {\bibinfo  {journal} {Phys.~Rev.~B}\ }\textbf {\bibinfo
  {volume} {53}},\ \bibinfo {pages} {3775} (\bibinfo {year}
  {1996})}\BibitemShut {NoStop}%
\bibitem [{\citenamefont {Becke}(1993)}]{Becke}%
  \BibitemOpen
  \bibfield  {author} {\bibinfo {author} {\bibfnamefont {A.~D.}\ \bibnamefont
  {Becke}},\ }\href {\doibase 10.1063/1.464913} {\bibfield  {journal} {\bibinfo
   {journal} {j. chem. phys.}\ }\textbf {\bibinfo {volume} {98}},\ \bibinfo
  {pages} {5648} (\bibinfo {year} {1993})}\BibitemShut {NoStop}%
\bibitem [{\citenamefont {Valiev}\ \emph {et~al.}(2010)\citenamefont {Valiev},
  \citenamefont {Bylaska}, \citenamefont {Govind}, \citenamefont {Kowalski},
  \citenamefont {Straatsma}, \citenamefont {van Dam}, \citenamefont {Wang},
  \citenamefont {Nieplocha}, \citenamefont {Apra}, \citenamefont {Windus},\
  and\ \citenamefont {de~Jong}}]{nwchem}%
  \BibitemOpen
  \bibfield  {author} {\bibinfo {author} {\bibfnamefont {M.}~\bibnamefont
  {Valiev}}, \bibinfo {author} {\bibfnamefont {E.}~\bibnamefont {Bylaska}},
  \bibinfo {author} {\bibfnamefont {N.}~\bibnamefont {Govind}}, \bibinfo
  {author} {\bibfnamefont {K.}~\bibnamefont {Kowalski}}, \bibinfo {author}
  {\bibfnamefont {T.}~\bibnamefont {Straatsma}}, \bibinfo {author}
  {\bibfnamefont {H.}~\bibnamefont {van Dam}}, \bibinfo {author} {\bibfnamefont
  {D.}~\bibnamefont {Wang}}, \bibinfo {author} {\bibfnamefont {J.}~\bibnamefont
  {Nieplocha}}, \bibinfo {author} {\bibfnamefont {E.}~\bibnamefont {Apra}},
  \bibinfo {author} {\bibfnamefont {T.}~\bibnamefont {Windus}}, \ and\ \bibinfo
  {author} {\bibfnamefont {W.}~\bibnamefont {de~Jong}},\ }\href {\doibase
  10.1016/j.cpc.2010.04.018} {\bibfield  {journal} {\bibinfo  {journal}
  {Comput. Phys. Commun.}\ }\textbf {\bibinfo {volume} {181}},\ \bibinfo
  {pages} {1477} (\bibinfo {year} {2010})}\BibitemShut {NoStop}%
\bibitem [{\citenamefont {Brouet}\ \emph {et~al.}(2001)\citenamefont {Brouet},
  \citenamefont {Alloul}, \citenamefont {Le}, \citenamefont {Garaj},\ and\
  \citenamefont {Forr\'o}}]{brouet01}%
  \BibitemOpen
  \bibfield  {author} {\bibinfo {author} {\bibfnamefont {V.}~\bibnamefont
  {Brouet}}, \bibinfo {author} {\bibfnamefont {H.}~\bibnamefont {Alloul}},
  \bibinfo {author} {\bibfnamefont {T.-N.}\ \bibnamefont {Le}}, \bibinfo
  {author} {\bibfnamefont {S.}~\bibnamefont {Garaj}}, \ and\ \bibinfo {author}
  {\bibfnamefont {L.}~\bibnamefont {Forr\'o}},\ }\href {\doibase
  10.1103/PhysRevLett.86.4680} {\bibfield  {journal} {\bibinfo  {journal}
  {Phys. Rev. Lett.}\ }\textbf {\bibinfo {volume} {86}},\ \bibinfo {pages}
  {4680} (\bibinfo {year} {2001})}\BibitemShut {NoStop}%
\bibitem [{\citenamefont {Manini}\ \emph {et~al.}(1994)\citenamefont {Manini},
  \citenamefont {Tosatti},\ and\ \citenamefont {Auerbach}}]{manini94}%
  \BibitemOpen
  \bibfield  {author} {\bibinfo {author} {\bibfnamefont {N.}~\bibnamefont
  {Manini}}, \bibinfo {author} {\bibfnamefont {E.}~\bibnamefont {Tosatti}}, \
  and\ \bibinfo {author} {\bibfnamefont {A.}~\bibnamefont {Auerbach}},\ }\href
  {\doibase 10.1103/PhysRevB.49.13008} {\bibfield  {journal} {\bibinfo
  {journal} {Phys. Rev. B}\ }\textbf {\bibinfo {volume} {49}},\ \bibinfo
  {pages} {13008} (\bibinfo {year} {1994})}\BibitemShut {NoStop}%
\bibitem [{\citenamefont {Capone}\ \emph {et~al.}(2004)\citenamefont {Capone},
  \citenamefont {Fabrizio}, \citenamefont {Castellani},\ and\ \citenamefont
  {Tosatti}}]{Capone04}%
  \BibitemOpen
  \bibfield  {author} {\bibinfo {author} {\bibfnamefont {M.}~\bibnamefont
  {Capone}}, \bibinfo {author} {\bibfnamefont {M.}~\bibnamefont {Fabrizio}},
  \bibinfo {author} {\bibfnamefont {C.}~\bibnamefont {Castellani}}, \ and\
  \bibinfo {author} {\bibfnamefont {E.}~\bibnamefont {Tosatti}},\ }\href
  {\doibase 10.1103/PhysRevLett.93.047001} {\bibfield  {journal} {\bibinfo
  {journal} {Phys. Rev. Lett.}\ }\textbf {\bibinfo {volume} {93}},\ \bibinfo
  {pages} {047001} (\bibinfo {year} {2004})}\BibitemShut {NoStop}%
\bibitem [{\citenamefont {Ishihara}\ \emph
  {et~al.}(1997{\natexlab{b}})\citenamefont {Ishihara}, \citenamefont {Inoue},\
  and\ \citenamefont {Maekawa}}]{Ishihara97}%
  \BibitemOpen
  \bibfield  {author} {\bibinfo {author} {\bibfnamefont {S.}~\bibnamefont
  {Ishihara}}, \bibinfo {author} {\bibfnamefont {J.}~\bibnamefont {Inoue}}, \
  and\ \bibinfo {author} {\bibfnamefont {S.}~\bibnamefont {Maekawa}},\ }\href
  {\doibase 10.1103/PhysRevB.55.8280} {\bibfield  {journal} {\bibinfo
  {journal} {Phys. Rev. B}\ }\textbf {\bibinfo {volume} {55}},\ \bibinfo
  {pages} {8280} (\bibinfo {year} {1997}{\natexlab{b}})}\BibitemShut {NoStop}%
\bibitem [{\citenamefont {van~den Brink}\ \emph {et~al.}(1998)\citenamefont
  {van~den Brink}, \citenamefont {Stekelenburg}, \citenamefont {Khomskii},
  \citenamefont {Sawatzky},\ and\ \citenamefont {Kugel}}]{vanderBrink98}%
  \BibitemOpen
  \bibfield  {author} {\bibinfo {author} {\bibfnamefont {J.}~\bibnamefont
  {van~den Brink}}, \bibinfo {author} {\bibfnamefont {W.}~\bibnamefont
  {Stekelenburg}}, \bibinfo {author} {\bibfnamefont {D.~I.}\ \bibnamefont
  {Khomskii}}, \bibinfo {author} {\bibfnamefont {G.~A.}\ \bibnamefont
  {Sawatzky}}, \ and\ \bibinfo {author} {\bibfnamefont {K.~I.}\ \bibnamefont
  {Kugel}},\ }\href {\doibase 10.1103/PhysRevB.58.10276} {\bibfield  {journal}
  {\bibinfo  {journal} {Phys. Rev. B}\ }\textbf {\bibinfo {volume} {58}},\
  \bibinfo {pages} {10276} (\bibinfo {year} {1998})}\BibitemShut {NoStop}%
\bibitem [{\citenamefont {Klupp}\ \emph {et~al.}(2012)\citenamefont {Klupp},
  \citenamefont {Matus}, \citenamefont {Kamar\'{a}s}, \citenamefont {Ganin},
  \citenamefont {{M}c{L}ennan}, \citenamefont {Rosseinsky}, \citenamefont
  {Takabayashi}, \citenamefont {{M}c{D}onald},\ and\ \citenamefont
  {Prassides}}]{klupp12}%
  \BibitemOpen
  \bibfield  {author} {\bibinfo {author} {\bibfnamefont {G.}~\bibnamefont
  {Klupp}}, \bibinfo {author} {\bibfnamefont {P.}~\bibnamefont {Matus}},
  \bibinfo {author} {\bibfnamefont {K.}~\bibnamefont {Kamar\'{a}s}}, \bibinfo
  {author} {\bibfnamefont {A.~Y.}\ \bibnamefont {Ganin}}, \bibinfo {author}
  {\bibfnamefont {A.}~\bibnamefont {{M}c{L}ennan}}, \bibinfo {author}
  {\bibfnamefont {M.~J.}\ \bibnamefont {Rosseinsky}}, \bibinfo {author}
  {\bibfnamefont {Y.}~\bibnamefont {Takabayashi}}, \bibinfo {author}
  {\bibfnamefont {M.~T.}\ \bibnamefont {{M}c{D}onald}}, \ and\ \bibinfo
  {author} {\bibfnamefont {K.}~\bibnamefont {Prassides}},\ }\href {\doibase
  10.1038/ncomms1910} {\bibfield  {journal} {\bibinfo  {journal} {Nature
  Communications}\ }\textbf {\bibinfo {volume} {3}},\ \bibinfo {pages} {912}
  (\bibinfo {year} {2012})}\BibitemShut {NoStop}%
\bibitem [{\citenamefont {Kamar\'{a}s}\ \emph {et~al.}(2013)\citenamefont
  {Kamar\'{a}s}, \citenamefont {Klupp}, \citenamefont {Matus}, \citenamefont
  {Ganin}, \citenamefont {{McL}ennan}, \citenamefont {Rosseinsky},
  \citenamefont {Takabayashi}, \citenamefont {{McD}onald},\ and\ \citenamefont
  {Prassides}}]{kamaras13}%
  \BibitemOpen
  \bibfield  {author} {\bibinfo {author} {\bibfnamefont {K.}~\bibnamefont
  {Kamar\'{a}s}}, \bibinfo {author} {\bibfnamefont {G.}~\bibnamefont {Klupp}},
  \bibinfo {author} {\bibfnamefont {P.}~\bibnamefont {Matus}}, \bibinfo
  {author} {\bibfnamefont {A.~Y.}\ \bibnamefont {Ganin}}, \bibinfo {author}
  {\bibfnamefont {A.}~\bibnamefont {{McL}ennan}}, \bibinfo {author}
  {\bibfnamefont {M.~J.}\ \bibnamefont {Rosseinsky}}, \bibinfo {author}
  {\bibfnamefont {Y.}~\bibnamefont {Takabayashi}}, \bibinfo {author}
  {\bibfnamefont {M.~T.}\ \bibnamefont {{McD}onald}}, \ and\ \bibinfo {author}
  {\bibfnamefont {K.}~\bibnamefont {Prassides}},\ }\href {\doibase
  10.1088/1742-6596/428/1/012002} {\bibfield  {journal} {\bibinfo  {journal}
  {J. Phys.: Conf. Ser.}\ }\textbf {\bibinfo {volume} {428}},\ \bibinfo {pages}
  {012002} (\bibinfo {year} {2013})}\BibitemShut {NoStop}%
\end{thebibliography}
\end{document}